\documentclass[12pt]{article}
\usepackage{graphicx}

\newcommand\pubnumber{}
\newcommand\pubdate{\today}

\def\institute{Institut f\"ur Experimentelle Kernphysik\\
Karlsruher Institut f\"ur Technologie, 76131 Karlsruhe, GERMANY}

\def\Title#1{\begin{center} {\Large #1 } \end{center}}
\def\Author#1{\begin{center}{ \sc #1} \end{center}}
\def\Address#1{\begin{center}{ \it #1} \end{center}}

\newcommand\pubblock{\rightline{\begin{tabular}{l} \pubnumber\\
         \pubdate  \end{tabular}}}
\newenvironment{Abstract}{\begin{quotation}  }{\end{quotation}}
\newenvironment{Presented}{\begin{quotation} \begin{center} 
             PRESENTED AT\end{center}\bigskip 
      \begin{center}\begin{large}}{\end{large}\end{center} \end{quotation}}

\def\beq{\begin{equation}}
\def\eeq#1{\label{#1}\end{equation}}
\def\eeqn{\end{equation}}

\def\beqa{\begin{eqnarray}}
\def\eeqa#1{\label{#1}\end{eqnarray}}
\def\eeqan{\end{eqnarray}}

\let\bar=\overbar

\def\Dslash{\not{\hbox{\kern-4pt $D$}}}
\def\dslash{\not{\hbox{\kern-2pt $\del$}}}

\def\msb{{\bar{\ssstyle M \kern -1pt S}}}

\newcommand{\sigmat}{\ensuremath{\sigma_{t\textrm{-ch.},\textrm{t}+\bar{\textrm{t}}}}}
\newcommand{\sigmattop}{\ensuremath{\sigma_{t\textrm{-ch.,t}}}}
\newcommand{\sigmatantitop}{\ensuremath{\sigma_{t\textrm{-ch.,}\bar{\textrm{t}}}}}
\newcommand{\xsectheotop}{ 136.02 }
\newcommand{\xsectheoantitop}{ 80.95 }
\newcommand{\xsectheo}{ 216.99 }
\renewcommand{\xsectheotop}{ 136.0 }
\renewcommand{\xsectheoantitop}{ 81.0 }
\renewcommand{\xsectheo}{ 217.0 }
\newcommand{\xsectheotopscale}{\ensuremath{^{+4.09}_{-2.92}}}
\newcommand{\xsectheoantitopscale}{\ensuremath{^{+2.53}_{-1.71}}}
\newcommand{\xsectheoscale}{\ensuremath{^{+6.62}_{-4.64}}}
\renewcommand{\xsectheotopscale}{\ensuremath{^{+4.1}_{-2.9}}}
\renewcommand{\xsectheoantitopscale}{\ensuremath{^{+2.5}_{-1.7}}}
\renewcommand{\xsectheoscale}{\ensuremath{^{+6.6}_{-4.6}}}
\newcommand{\xsectheotoppdf}{\ensuremath{\pm3.52}}
\newcommand{\xsectheoantitoppdf}{\ensuremath{\pm3.18}}
\newcommand{\xsectheopdf}{\ensuremath{\pm6.16}}
\renewcommand{\xsectheotoppdf}{\ensuremath{\pm3.5}}
\renewcommand{\xsectheoantitoppdf}{\ensuremath{\pm3.2}}
\renewcommand{\xsectheopdf}{\ensuremath{\pm6.2}}

\begin{document}
\begin{titlepage}
\pubblock

\vfill
\Title{Measurement of the $t$-channel single-top quark production cross section at 13 TeV with the CMS detector}
\vfill
\Author{Nils Faltermann \\on behalf of the CMS collaboration}
\Address{\institute}
\vfill
\begin{Abstract}
The electroweak production of single-top quarks in the $t$ channel can be changed by any deviation from the standard model, it is therefore an excellent opportunity to search for new physics. In this poster the recent cross-section measurement performed by the CMS collaboration is presented with the full 2015 dataset of the LHC Run II at a center-of-mass energy of 13 TeV. The cross section and the top/antitop ratio is extracted using a binned maximum likelihood fit to the distribution of a multivariate classifier in events containing one isolated muon in the final state.
\end{Abstract}
\vfill
\begin{Presented}
$9^{\mathrm{th}}$ International Workshop on Top Quark Physics\\
Olomouc, Czech Republic, September 19--23, 2016
\end{Presented}
\vfill
\end{titlepage}
\def\thefootnote{\fnsymbol{footnote}}
\setcounter{footnote}{0}
\section{Introduction}
\begin{figure}[htb]
\centering
\includegraphics[height=1.5in]{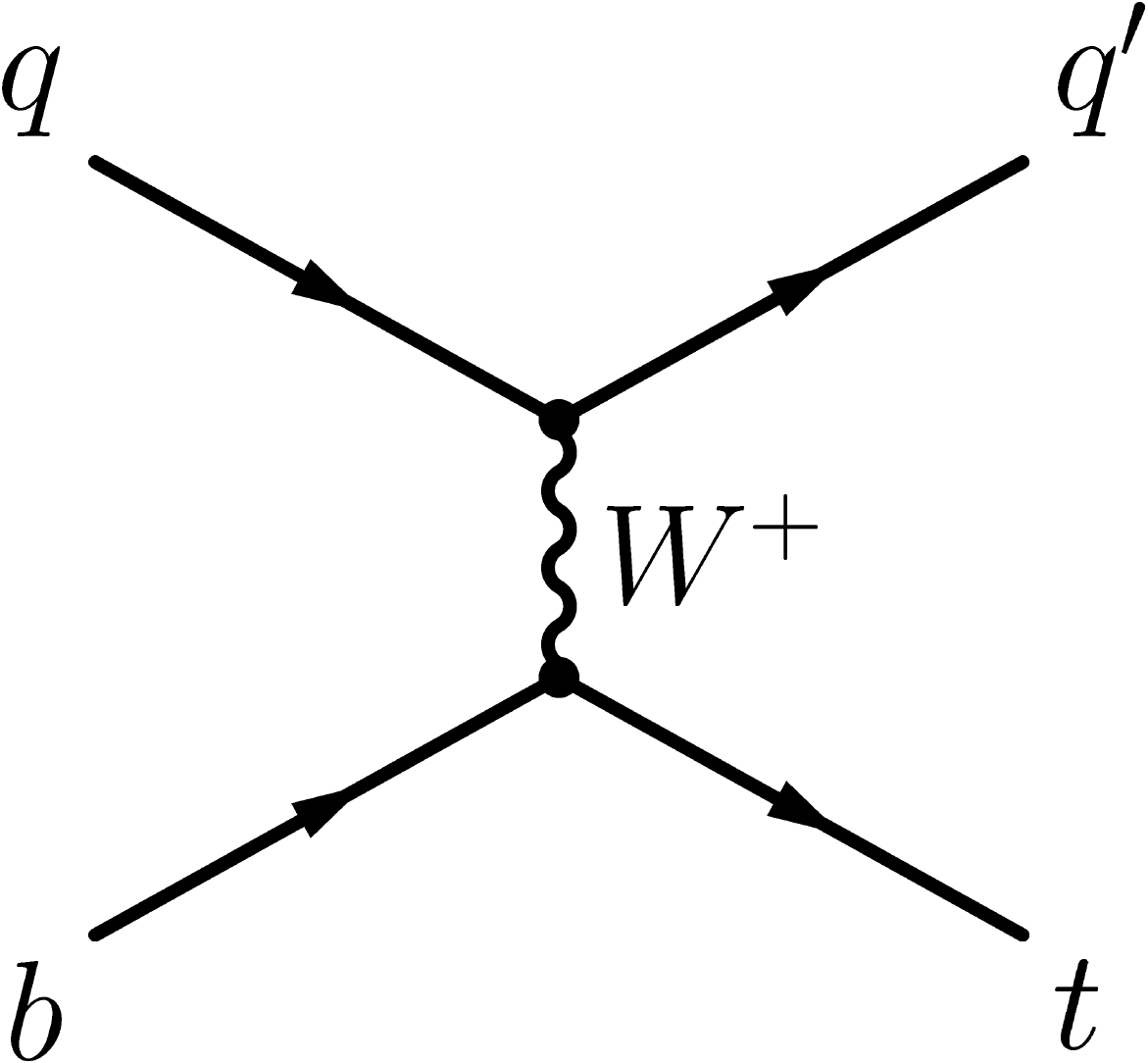}\qquad\qquad
\includegraphics[height=1.5in]{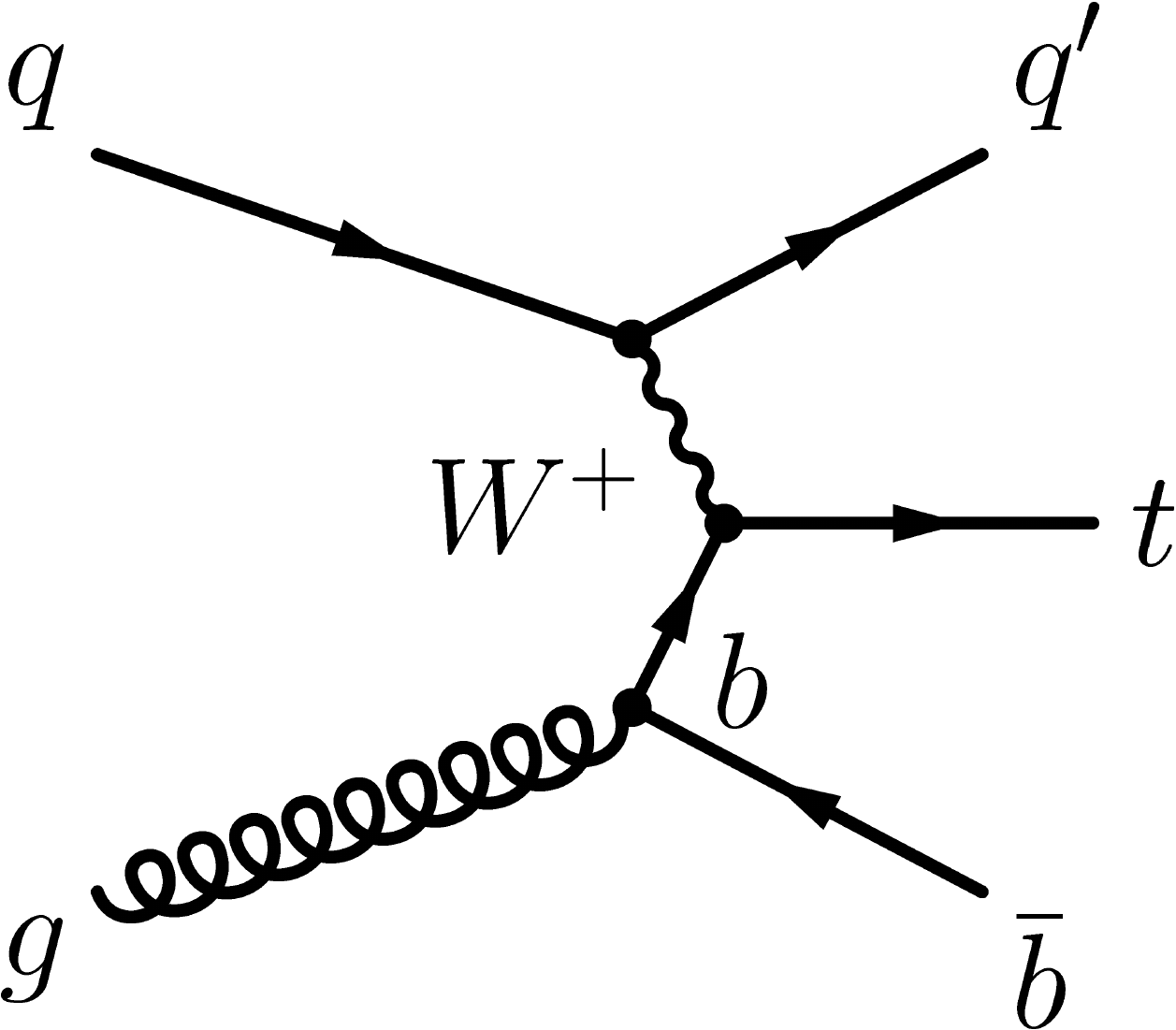}
\caption{Leading order Feynman diagrams for the $t$-channel production of single top quarks. In the left diagram the bottom quark comes directly from the initial proton (five flavor scheme), while in the right diagram a gluon needs to split into a bottom quark-antiquark pair first (four flavor scheme).}
\label{fig:fdia}
\end{figure}

At the LHC top quarks are mainly produced in top-antiquark pairs through the strong interaction. In addition, top quarks can also be produced as singly via the electroweak force. This is achieved through the interaction of a W boson and a bottom quark. About $70\%$ of all single top quarks are produced via the $t$ channel at a center-of-mass energy of $13\,\textrm{TeV}$. This process can either be described in the five flavor scheme, where the bottom quark is one of the initial partons ($2 \rightarrow 2$ process), and the four flavor scheme, where a gluon splits into a bottom quark-antiquark pair ($2 \rightarrow 3$ process). The leading order Feynman diagrams for the $t$-channel production in both schemes are shown in Figure~\ref{fig:fdia}. The remaining $30\%$ are achieved via tW associated production and the $s$-channel production.
The theoretical cross sections at $13\,\textrm{TeV}$ for top quark, antitop quark and inclusive $t$-channel production at next-to-leading order, calculated using {\sc HATHOR}~\cite{hathor}, are:
\begin{eqnarray*}
\sigmattop &=& \xsectheotop  \, \xsectheotopscale\,{\rm (scale)} \xsectheotoppdf \,(\textrm{PDF}{+}{\alpha_{\rm S}})\,\textrm{pb}, \\
\sigmatantitop &=&  \xsectheoantitop \, \xsectheoantitopscale\, {\rm (scale)} \xsectheoantitoppdf\,(\textrm{PDF}{+}{\alpha_{\rm S}})\,\textrm{pb}, \\
\sigmat &=&  \xsectheo \, \xsectheoscale\, {\rm (scale)} \xsectheopdf \,(\textrm{PDF}{+}{\alpha_{\rm S}})\,\textrm{pb}
\end{eqnarray*}
\section{Measurement}
The analyzed data consists of proton-proton collisions of the LHC recorded with the CMS experiment in 2015~\cite{cms}. This dataset corresponds to an integrated luminosity of $L\,=\, 2.3\,\textrm{fb}^{-1}$. The top quark decays in almost all cases into a W boson and a bottom quark, thus the top quark decay is characterized by the decay of the W boson. In this analysis, only decays of the W boson into a muon and a neutrino are considered, where the muon can either come directly from the W boson or through the decay of a tau lepton. In summary, the final state of the process consists of a light flavored quark recoiling against the top quark, usually in forward direction, a bottom quark from the top quark decay, a muon and a neutrino. A second bottom quark can also be present in the final state as shown in Figure~\ref{fig:fdia}, but it often fails the detector acceptance.\\

Events are selected which contain exactly two jets, from which only one is identified as originating from a bottom quark, and one isolated muon in the final state. The neutrino escapes the detector without any interaction, but it can be indirectly measured by the missing transverse energy of an event. By constraining the W boson mass to the literature value of $80.4\,\textrm{GeV}$, a quadratic equation can be obtained for the longitudinal momentum of the neutrino. The top quark can then be reconstructed by combining the momenta of the b tagged jet, the muon and the neutrino.\\

\begin{figure}[htb]
\centering
\includegraphics[height=1.8in]{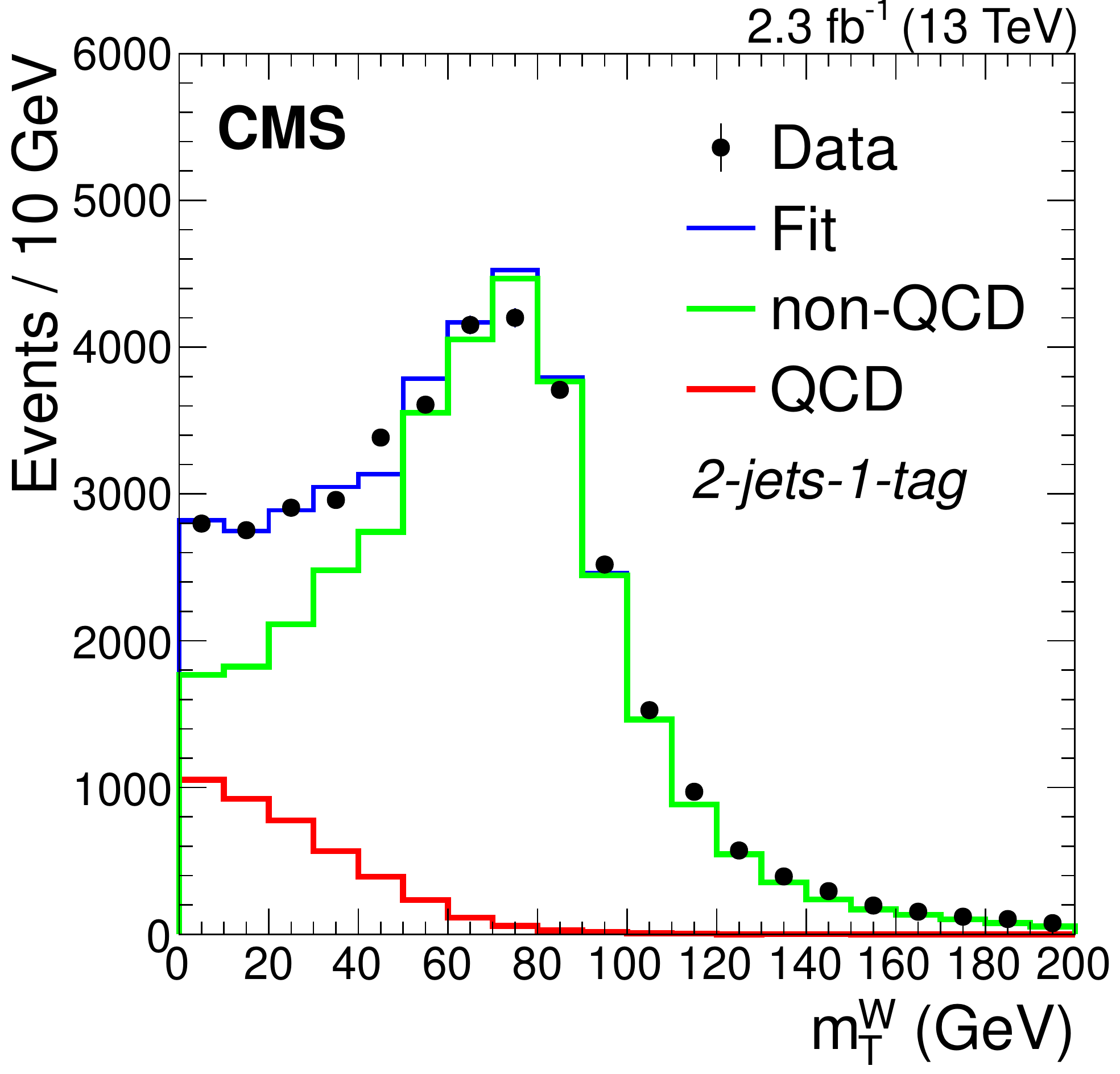}
\includegraphics[height=1.8in]{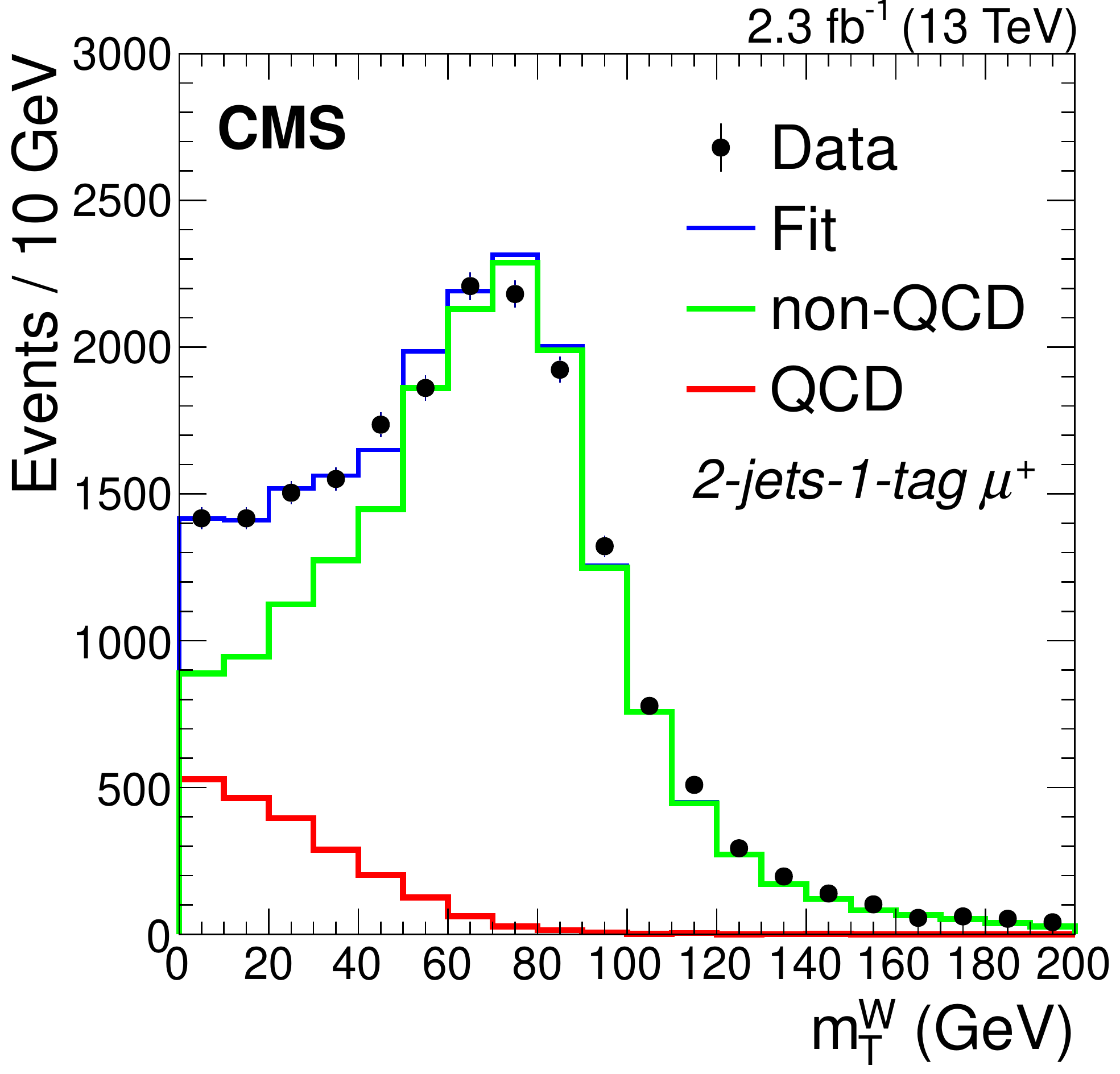}
\includegraphics[height=1.8in]{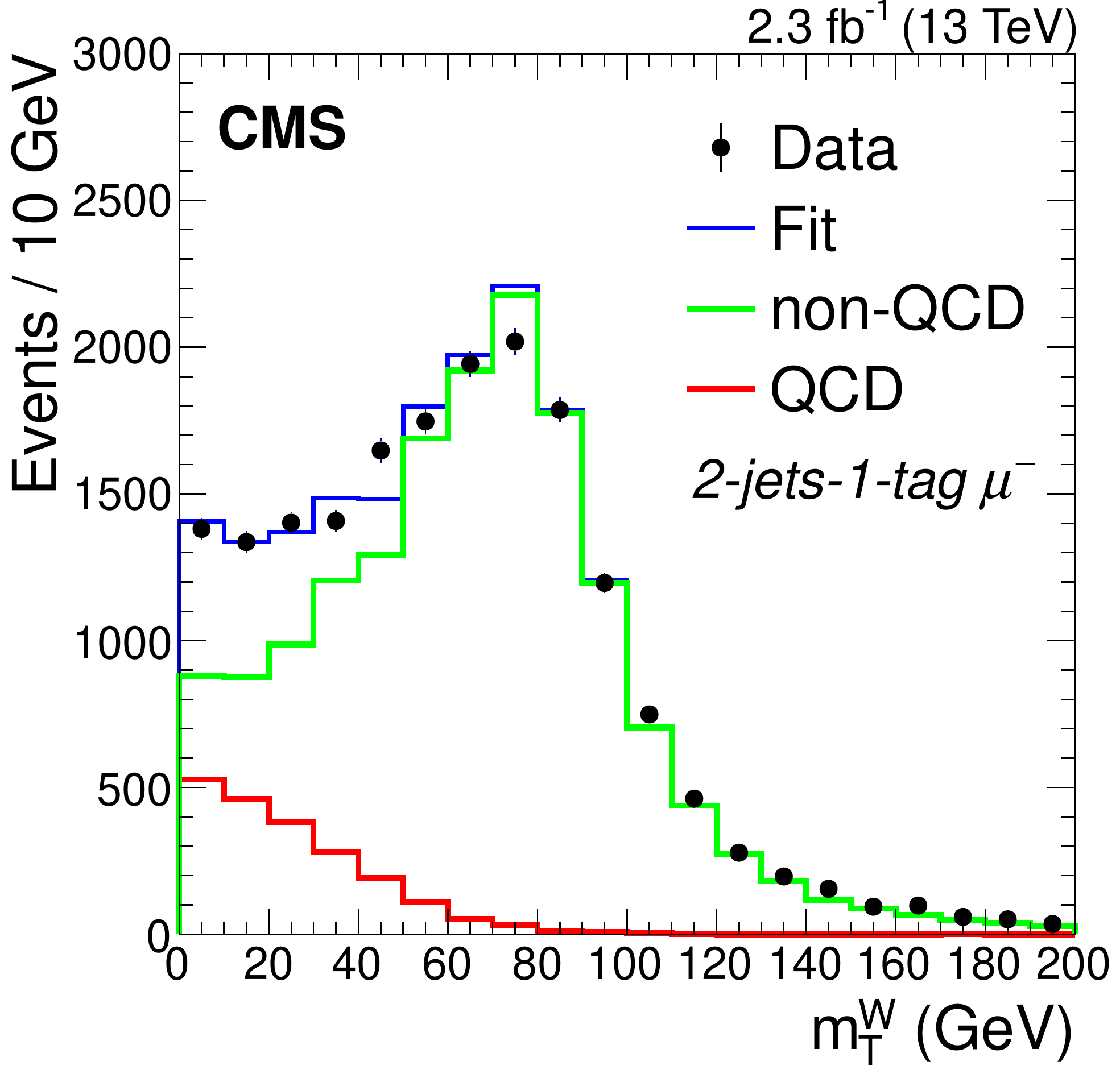}
\caption{Results of the QCD estimation for the inclusive selection (left), only positively charged muons (middle) and only negatively charged muons (right).}
\label{fig:qcd}
\end{figure}

Since the simulation of the QCD multijet background is not reliable a data-driven estimation is applied. The distribution of the transverse W boson mass $m_{\mathrm{T}}$ is different for QCD and non-QCD processes, therefore a fit to this distribution is performed. The result is shown in Figure~\ref{fig:qcd}. To further suppress this contribution a cut on $m_{\mathrm{T}}\,>\,50\,\mathrm{GeV}$ is applied.
All other processes are estimated with Monte Carlo simulations. The expected and observed event yields in the signal region are shown in Table~\ref{tab:yields}.
\begin{table}[t]
\begin{center}
\begin{tabular}{ c|c|c } 
Process & $\mu^{+}$& $\mu^{-}$ \\
\hline
Top quark (t$\bar{\textrm{t}}$~and tW) & $7048 \pm13$ & $7056 \pm 13$ \\
W+jets and Z+jets & $2837 \pm 83 $ & $ 2564 \pm 77 $ \\
QCD multijet &$ 302 \pm151$ & $ 262 \pm 131$ \\
\hline
Single top quark $t$-channel&$1539 \pm 13 $ & $ 977 \pm 10 $ \\
\hline
Total expected& $11726 \pm 173$ &$ 10859 \pm 153 $\\
\hline
Data&11877&  11017  \\
\end{tabular} 
\caption{Expected and observed yields for events with positively and negatively charged muons. The expected yields are taken from Monte Carlo simulations and the uncertainty is corresponding to the sample size, with the exception of the QCD multijet background which is estimated from data with a $50\%$ uncertainty.}
\label{tab:yields}
\end{center}
\end{table}\\

\begin{figure}[htb]
\centering
\includegraphics[height=1.8in]{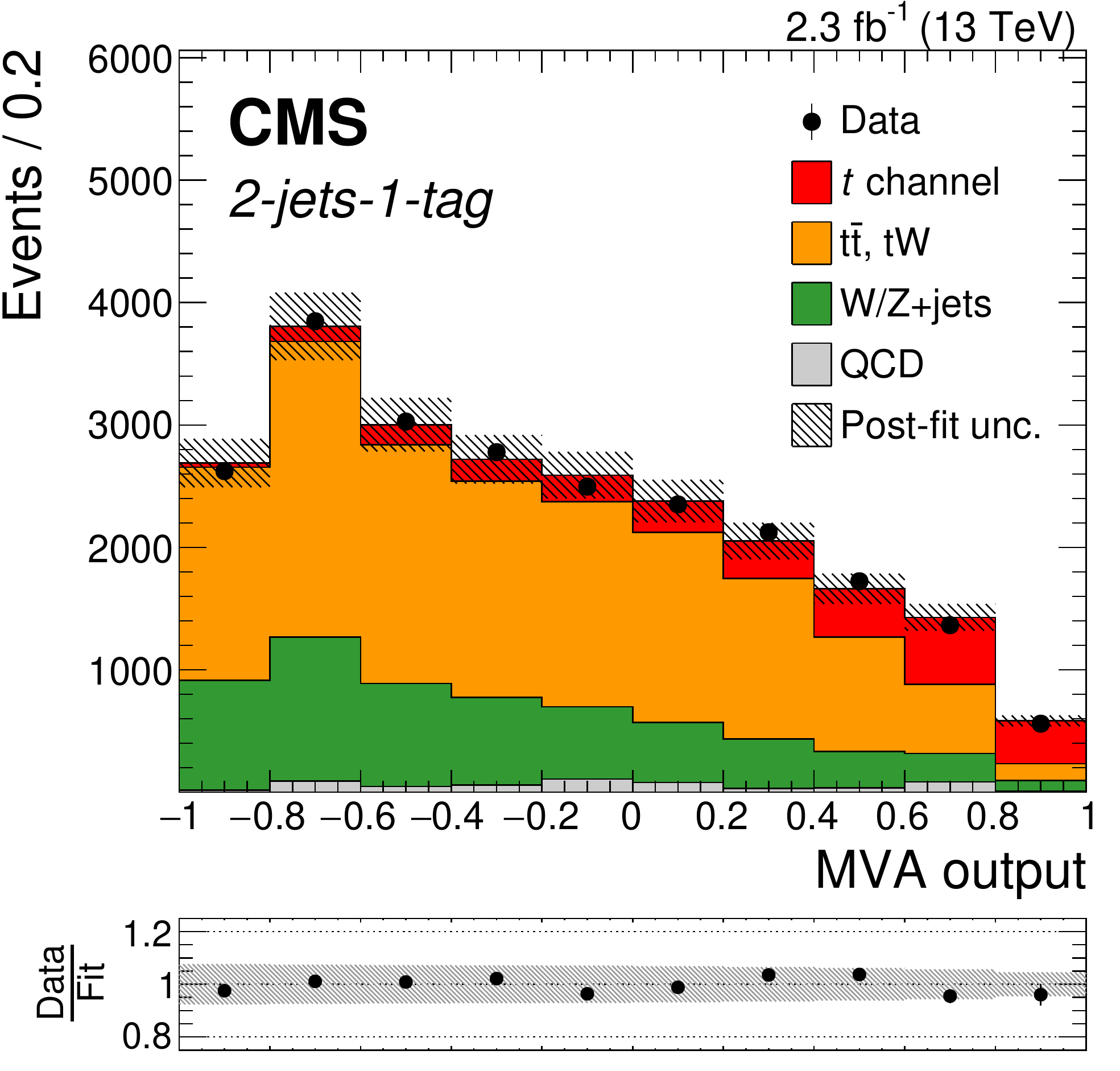}
\includegraphics[height=1.8in]{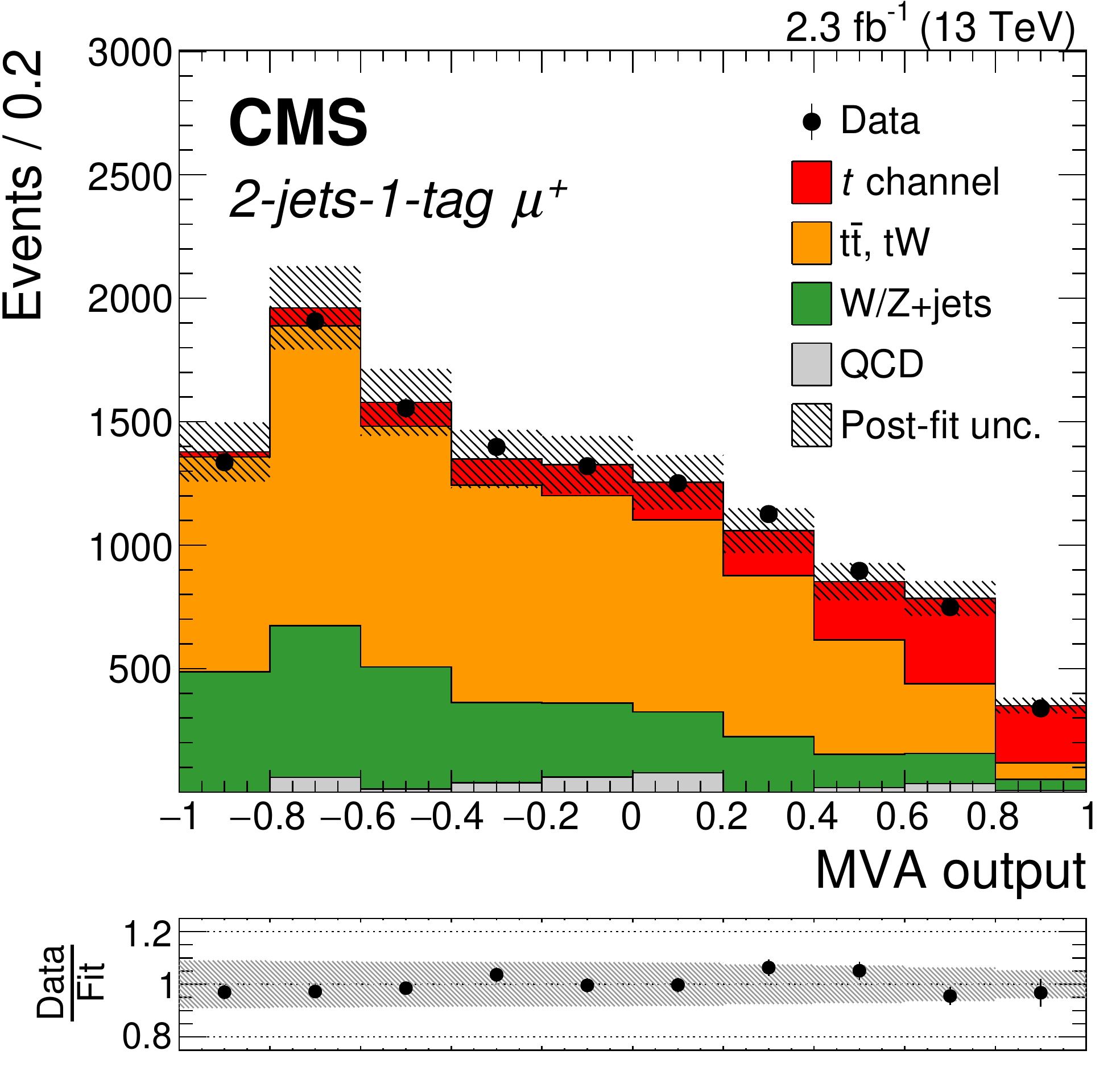}
\includegraphics[height=1.8in]{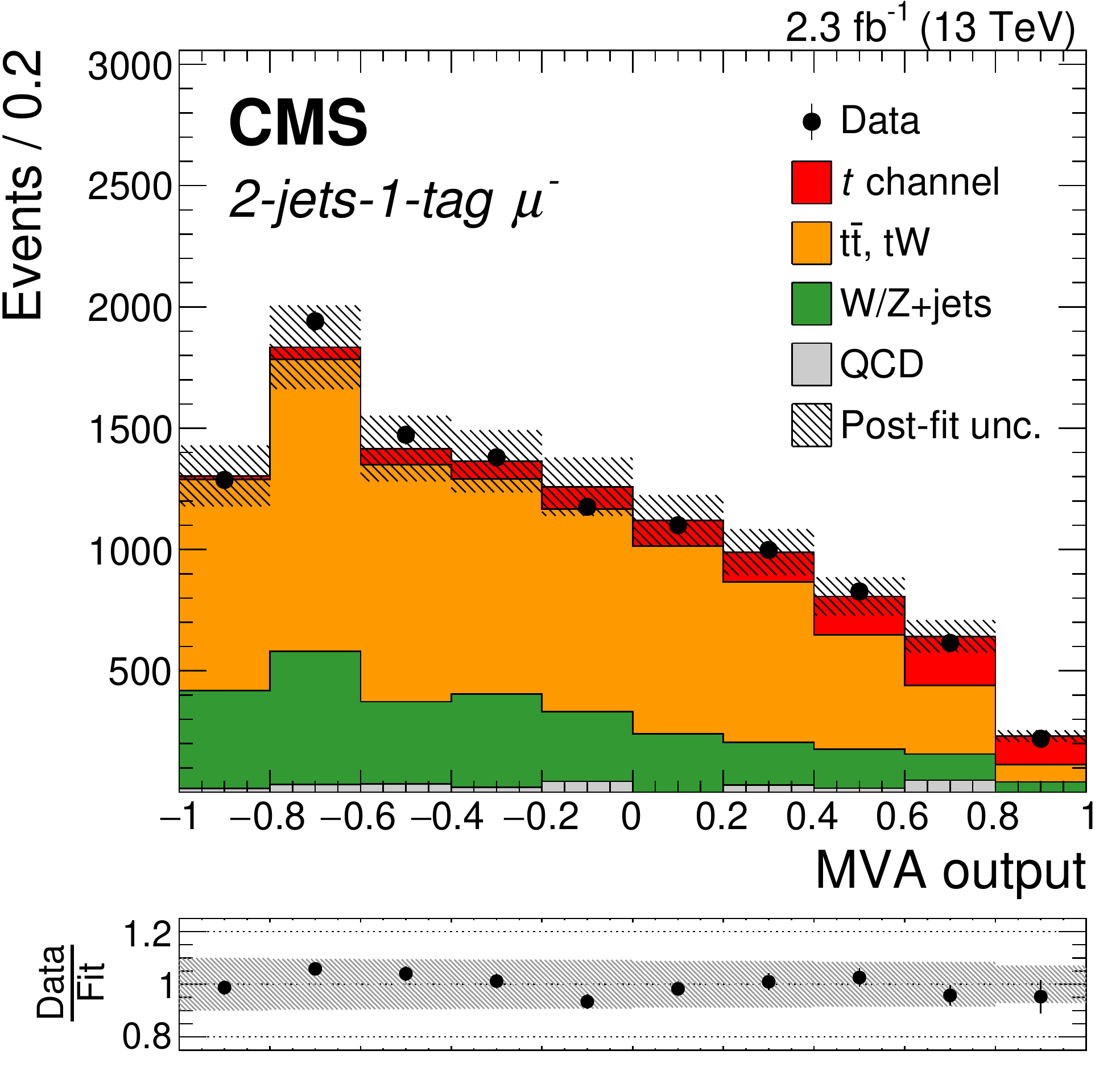}
\caption{MVA distribution in the 2-jets--1-tag signal region for the inclusive selection (left), only positively charged muons (middle) and only negatively charged muons (right). The simulation is scaled to the result of the fit. A good agreement of expectation and data within the uncertainties is observed.}
\label{fig:sr}
\end{figure}

To distinguish the signal process from the different background processes a neural network is trained with eleven kinematic variables, such as the pseudorapidity of the recoiling quark or the reconstructed top quark mass. A maximum likelihood fit is then performed to the output distribution of the neural network, where the signal is unconstrained and all background processes have a specific prior with a width according to the theoretical uncertainty of their cross section. The fit is simultaneously done in the signal region and two control regions to constrain the top quark background. The distribution in the signal region is shown in Figure~\ref{fig:sr}.
\section{Results}
\begin{figure}[htb]
\centering
\includegraphics[height=3in]{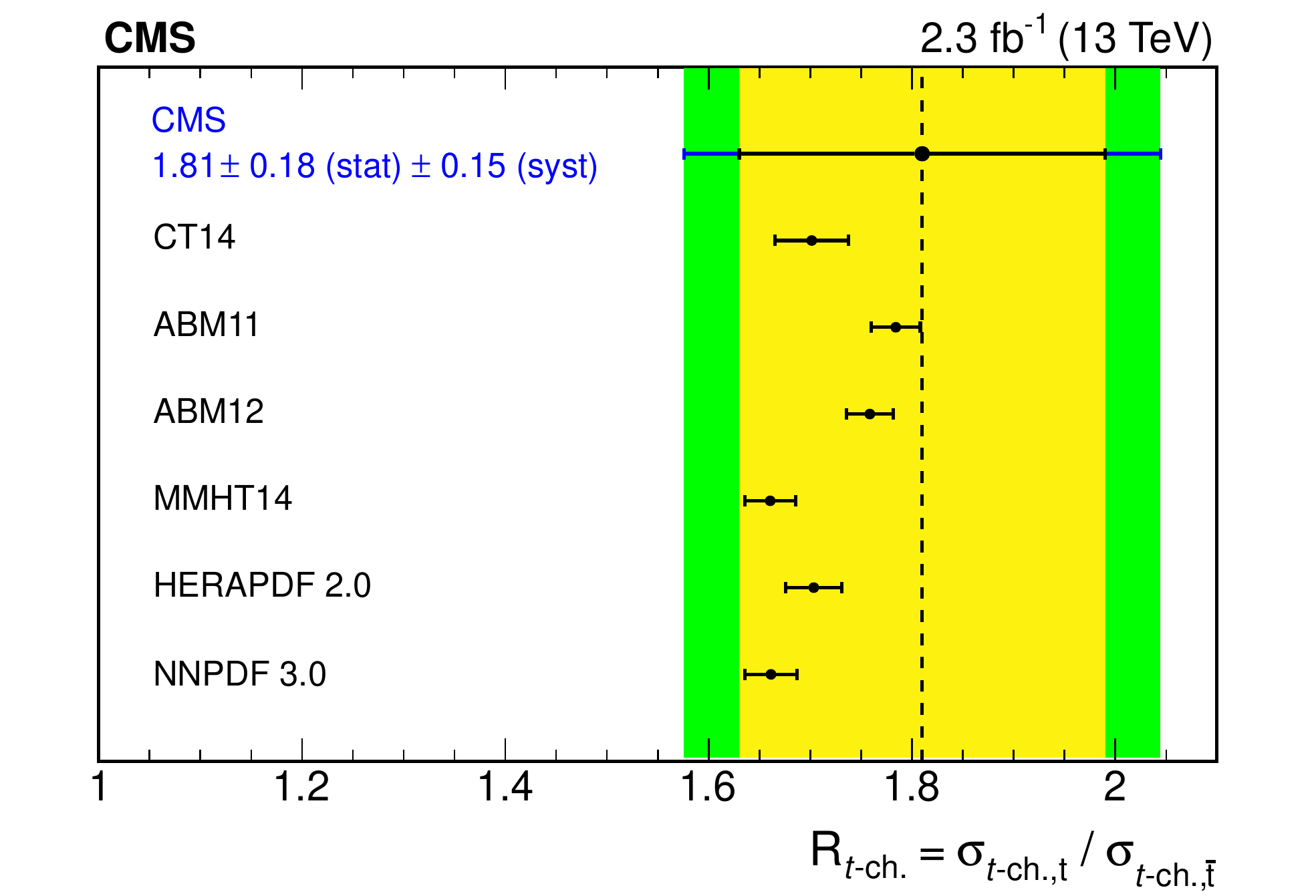}
\caption{The measured ratio of top and antitop production cross section in comparison with the prediction from different PDF sets. The uncertainties on the PDF sets include uncertainties on the factorization and renormalization scales, uncertainty on the top quark mass and statistical uncertainty. For the CMS result the NNPDF 3.0 PDF set was used.}
\label{fig:ratio}
\end{figure}

The measured cross sections for top quark, antitop quark and inclusive $t$ channel production are:
\begin{eqnarray*}
\sigma_{t\textrm{-ch.,t}} & = & 150 \pm 8\,\rm{(stat)} \pm 9\,\rm{(exp)} \pm 18\,\rm{(theo)}  \pm 4\,\rm{(lumi)}\,\rm{pb}\\
\sigma_{t\textrm{-ch.,}\bar{\textrm{t}}} & = & 82 \pm 10\,\rm{(stat)} \pm 4\,\rm{(exp)} \pm 11\,\rm{(theo)}  \pm 2\,\rm{(lumi)}\,\rm{pb} \\
\sigma_{t\textrm{-ch.},\textrm{t}+\bar{\textrm{t}}} & = & 232 \pm 13\,\rm{(stat)} \pm 12\,\rm{(exp)} \pm 26\,\rm{(theo)} \pm 6\,\rm{(lumi)}\,\rm{pb} 
\end{eqnarray*}
The ratio of top quark and antitop quark cross section is sensitive to the ratio of up- and down-type quarks in the initial state and thus to the quark PDF. A comparison with different PDF sets is shown in Figure~\ref{fig:ratio}. All measured quantities are in agreement with the standard model predictions. Further details can be found in Ref.~\cite{paper}.

\end{document}